\documentclass[twocolumn,           % Format : preprint, twocolumn
               showpacs,            % Pacs : showpacs, noshowpacs
               nopreprintnumbers,     % Preprint: preprintnumbers,
               aps,                 % Society: ...
               prd,          	    % Journal Style : pra, prb, prc, prd, pre,
               letterpaper,             % Size : a4paper, ...
               groupeaddress,      % Affiliation (Title) : groupedaddress,
               nofootinbib,         % Footnote: footinbib, nofootinbib
               tightenlines,        % Remove additional spaces in a line
               floats,floatfix      % Floating pictures and tables
               ]{revtex4-1}

\usepackage{graphicx}% Include figure files
\usepackage{fancyhdr}
\usepackage[english]{babel}
\begin{document}

\pagestyle{fancy}
\thispagestyle{plain}
\fancypagestyle{plain}{

%%%HEADER%%%
\fancyhead[C]{\includegraphics[width=18.5cm]{head_foot/header_bar}}
\fancyhead[L]{\hspace{0cm}\vspace{1.5cm}\includegraphics[height=30pt]{head_foot/journal_name}}
\fancyhead[R]{\hspace{0cm}\vspace{1.7cm}\includegraphics[height=55pt]{head_foot/RSC_LOGO_CMYK}}
\renewcommand{\headrulewidth}{0pt}
}
%%%END OF HEADER%%%

%%%MAIN TEXT%%%%

\title{
Size, vacancy and temperature effects on Young's modulus of silicene nanoribbons
}

\author{M.R. Ch\'avez-Castillo$^{1,2}$}
%\email{aspeitia@fisica.uaz.edu.mx}

\author{Mario A. Rodr\'iguez-Meza$^{2}$}
\email{marioalberto.rodriguez@inin.gob.mx}

\author{L. Meza-Montes$^{1}$}
\email{lilia@ifuap.buap.mx}

\affiliation{$^{1}$Instituto de F\'isica, Benem\'erita Universidad Aut\'onoma de Puebla, Apdo. Postal J-48, 72570, Puebla, Pue., M\'exico.
}

\affiliation{$^{2}$Departamento de F\'\i sica, Instituto Nacional de Investigaciones Nucleares, Apdo. Postal 18-1027, M\'exico D.F. 11801, M\'exico}

\begin{abstract}
We report results on the Young's modulus (YM) of defect-free and defective silicene nanoribbons (SNRs) as a function of length and temperature. In this study, we perform molecular dynamics simulations using the Environment-Dependent Interatomic Potential (EDIP) to describe the interaction of the Si atoms.  We find that the Young's modulus of pristine and defective SNRs increases with the ribbon length in both chirality directions. It is shown that the Young's modulus of defective SNRs exhibit a complex dependence on the combinations of vacancies. With respect to temperature, we find that YM for SNRs with and without vacancy defects shows a nonlinear behavior and it could be tailoring for a given length and chirality.
\end{abstract}

\keywords{Young's modulus, silicene, graphene, nanoribbons, two-dimensional materials}
\draft
\pacs{}
\date{\today}
\maketitle

\section{Introduction}

The beginning of the 21st century introduced a new class of materials called two-dimensional materials \cite{Novoselov2005, Miro2014}. The first identified material in this new category was graphene \cite{Novoselov2004}, which is a single atomic layer of carbon with a honeycomb structure. Since its discovery, in 2004, several studies have revealed the extraordinary properties of graphene, making it one of the most promising materials for applications in electronics \cite{Woszczyna2014}, optics \cite{Gunho2012}, sensors \cite{Wang2014}, energy storage \cite{Zhu2014}, water purification \cite{Han2013}, and biodevices \cite{Hu2015}. Besides the graphene, a large variety of these layered materials have been studied such as boron nitride \cite{Song2010, Novoselov2005, Sahin2009}, dichalcogenides \cite{Chhowalla2013, Wang2012, Butler2013, Novoselov2005, Miro2014, Sahin2009}, germanene \cite{Garcia2011, Balendhran2014, Cahangirov2009, Sahin2009, Lebegue2009, Liu2011,Davila2014}, silicene \cite{Vogt2012, Garcia2011,Balendhran2014,Cahangirov2009,Sahin2009,Lebegue2009,Liu2011,Lin2012,Houssa2015,Deepthi2014}, stanene \cite{Garcia2011, Xu2013, Balendhran2014}, phosphorene \cite{Liu2014, Li2014, Balendhran2014,Xia2014}, borophene \cite{Piazza2014,Zhai2014}, and most recently titanium trisulfide \cite{Zeng2015},  arsenene and antimonene \cite{Zhang2015}.  Of this growing family of 2D materials, silicene has emerged to replace, not only graphene, but also bulk silicon in the current electronics, because studies have shown that silicene has electronic properties similar to those of graphene \cite{Cahangirov2009, Lebegue2009}.  Silicene, analogous to graphene, is made of Si atoms arranged in a honeycomb structure. Nevertheless, it differs from graphene in the degree of flatness. While graphene has a perfectly flat stable structure,  silicene shows a buckled stable structure. It should be noted that silicene is a material whose structure has not been obtained yet, like graphene. So,  the existence of freestanding silicene sheets are so far hypothetical \cite{Cahangirov2009}. Despite this, research shows that silicene can be synthesized on ordered substrates like Ag \cite{Vogt2012, Chun2012, Enriquez2012, Jamgotchian2012, Chen2012, Chen2013},  ZrBr$_2$ \cite{Fleurence2012} and Ir \cite{Meng2013}.

Nowadays, the application of these layered materials in nanoelectromechanical systems (NEMS) has attracted much attention, so it is essential to understand their mechanical properties in order to exploit their potential.
Thus far, the outstanding electronic \cite{Novoselov2007, Castro2009} and mechanical \cite{Lee2008} properties of graphene, make it one of the most studied materials for such applications \cite{Standley2008, Garcia2008, Chen2009, Suk2014, Xining2015, Wang2015}. In this context, there are quite a few studies of the most fundamental mechanical property, which is the {\it Young's modulus} (YM) or {\it elastic modulus} defined as the ratio of the stress to the strain as a material is stretched. Experimentally the YM of graphene has been obtained for monolayer, bilayer and multiple layers. Reported values for a monolayer are 1.2 TPa \cite{Lee2008},  0.89 TPa \cite{Yupeng2012} and 2.4 TPa \cite{Lee2012} ; for a bilayer are 0.39 TPa \cite{Yupeng2012} and 2.0 TPa \cite{Lee2012}, and finally for a multilayer (less than 5 layers) is 0.5 TPa \cite{Frank2007}. From the theoretical point of view, not only density functional theory (DFT) and tight binding (TB) calculations have been made, but also molecular dynamics (MD)  and molecular mechanics (MM).  Employing different kinds of potentials, these methods have been used to determine the YM of graphene, with reported values of  {\it e. g.}  1.05 \cite{Hajgato2012} , 0.91 TPa \cite{Zhao2009}, 1.01 and 1.19 TPa, respectively \cite{Zhao2009, Lixin2013}. These values are in good agreement with those reported by Lee {\it et al.} \cite{Lee2008}. However, the TB calculation shows a slightly lower value than others studies.
Because NEMS and their components (nanostructures) have structural dimensions around or below 100 nm, graphene nanoribbons (GNRs) are good candidates to develop devices based on them. Thus, research on its mechanical behavior is of increasing interest. Results have shown that YM depends on the ribbons width as well as their chirality \cite{Xu2009, Zhao2009, Lu2011}.
Since silicene could be equivalent to graphene, with the advantage of an easy integration to the present technology, the interest of using silicene nanoribbons (SNRs) has opened a wide field of study. By now, several studies of  the mechanical properties  of pristine silicene \cite{Sahin2009, Ru2012, Zhao2012, Peng2013, Roman2014,Pei2014}  and SNRs  \cite{Topsakal2010, Jing2013, Ansari2014}  have been reported.

 A very important mechanical property in this kind of materials is the YM  because many design problems in nanoscale mechanical devices, like NEMS,  are related to it. Mechanical  properties of nanomaterials, as it is well known, exhibit a size dependence. YM could also be affected by the presence of defects in the nanomaterial. In honeycomb structures, such as graphene and silicene, vacancy defects are the main defects observed \cite{Ozcccelik2013}.  So,  a good understanding of the behavior of YM in nanomaterials with and without vacancy defects is helpful for design and fabrication of  reliable nanoscale mechanical devices.

In this paper, we apply classical molecular dynamics simulations to investigate the size and temperature dependence of the YM of free-standing SNRs. YM of pristine and defective SNRs as a function of length and temperature are reported. We found that the mechanical properties for pristine and defective SNRs exhibit a size dependence in both chirality directions. YM shows not only a size dependence but also a meaningful dependence of temperature and combinations of vacancy defects. The rest of the paper is organized as follows. In Section \ref{sec:Method}, we briefly describe our calculation method. In Section \ref{sec:Results}, the results are analyzed and finally, conclusions are given in Section \ref{sec:Concl}.

\section{Methodology}
\label{sec:Method}

To study the YM of low-buckled silicene nanoribbons \cite{Ansari2014, Vogt2012}, we perform molecular dynamics simulations of SNRs under uniaxial tension using LAMMPS \cite{Plimpton1995}. Si atoms interact via EDIP  \cite{Justo1998} potential as implemented in LAMMPS. All simulations are performed at room temperature for three squared-shaped SNRs, with and without defects, whose lengths are shown in Table ~\ref{tbl:sizes}. As the chirality matches with the direction of deformation, each ribbon length is represented by its chirality as ach or zz, and number labels size of the ribbon. In what follows, this nomenclature will be used to specify chirality direction and size. An {\it NVE} ensemble is used with 30 \AA\ vacuum space on each side along {\it z}-direction to relax the ideal structure (see Fig. ~\ref{fgr:YoungModulusSiliceneFig1} ). Fig.~\ref{SI-fgr:snapshotsrelax} 
in the Supplementary Information (SI) 
in the Appendix (hereafter named as SI),
shows  front and lateral views of a relaxed structure, where can be seen that the structure presents a variable buckling. Similar behavior is found in all structures considered in this study. After relaxing the structure, {\it NVT} simulations are performed with a time step of 1 fs. We apply periodic boundary conditions (PBC) along the direction of tension, which is also the chirality direction, as mentioned before. PBC are needed in order to avoid any boundary effect in the structure along the tensile direction. Perpendicular edges to the tensile direction are free, see Fig. \ref{fgr:YoungModulusSiliceneFig1}.  A constant strain rate of 0.005 ps$^{-1}$ is applied to the structure to simulate uniaxial tension every time step during 90 ps. For integration of the motion equations we use the velocity-Verlet algorithm with an integration time of 1 fs. For each simulation, we obtain a stress-strain curve, since the YM is generally defined as the corresponding slope of the linear regime (elastic region) curve. A linear interpolation is performed in order to find YM \cite{Lu2011}.

\begin{table}[h]
\small
  \caption{\ Armchair and zig-zag square-shaped sizes of  SNRs.  Chirality is represented by ach or zz and number labels size of the ribbon}
  \label{tbl:sizes}
  \begin{tabular*}{0.5\textwidth}{@{\extracolsep{\fill}}lll}
    \hline
    Length/\AA & \multicolumn{2}{c}{Chirality} \\
    & armchair & zig-zag\\
    \hline
    $32.48 \times 32.15$ & ach1 & zz1 \\
    $60.32 \times 60.28$ & ach2 & zz2 \\
    $81.20 \times 80.36$ & ach3 & zz3\\
    \hline
  \end{tabular*}
\end{table}

\begin{figure}[h]
\centering
  \includegraphics[height=8cm]{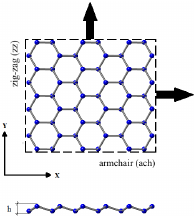}
  \caption{ Top: Ideal defect-free silicene, along with the definitions of  armchair and zig-zag directions, to illustrate  the simulation procedure. The dashed box indicates the boundaries of the simulation box. Arrows indicate the armchair or zig-zag straining direction. Bottom:  Side view of silicene structure with buckling distance $h=0.75$ \AA\cite{Ansari2014, Vogt2012}}
  \label{fgr:YoungModulusSiliceneFig1}
\end{figure}

\section{Results}
\label{sec:Results}

First, we perform simulations at room temperature. A bulk (sheet) silicene with 4050 Si atoms is considered. Young's  modulus obtained for this sheet are 154.7 and 152.5 GPa for zz and ach directions, respectively. These values are in good agreement with 154.8 and 153.8 GPa, obtained by Jing {\it et al.} \cite{Jing2013}.
In order to study the effect of different vacancies, simulations are performed for different lengths of NRs, with and without vacancies. Later on, temperature dependence is analyzed  (Sec. 3.2). 

\subsection{Size effects}

\subsubsection{Pristine nanoribbons.~~} 
\label{sec:pristinenrs}

\begin{figure*}
\centering
  \includegraphics[height=5cm]{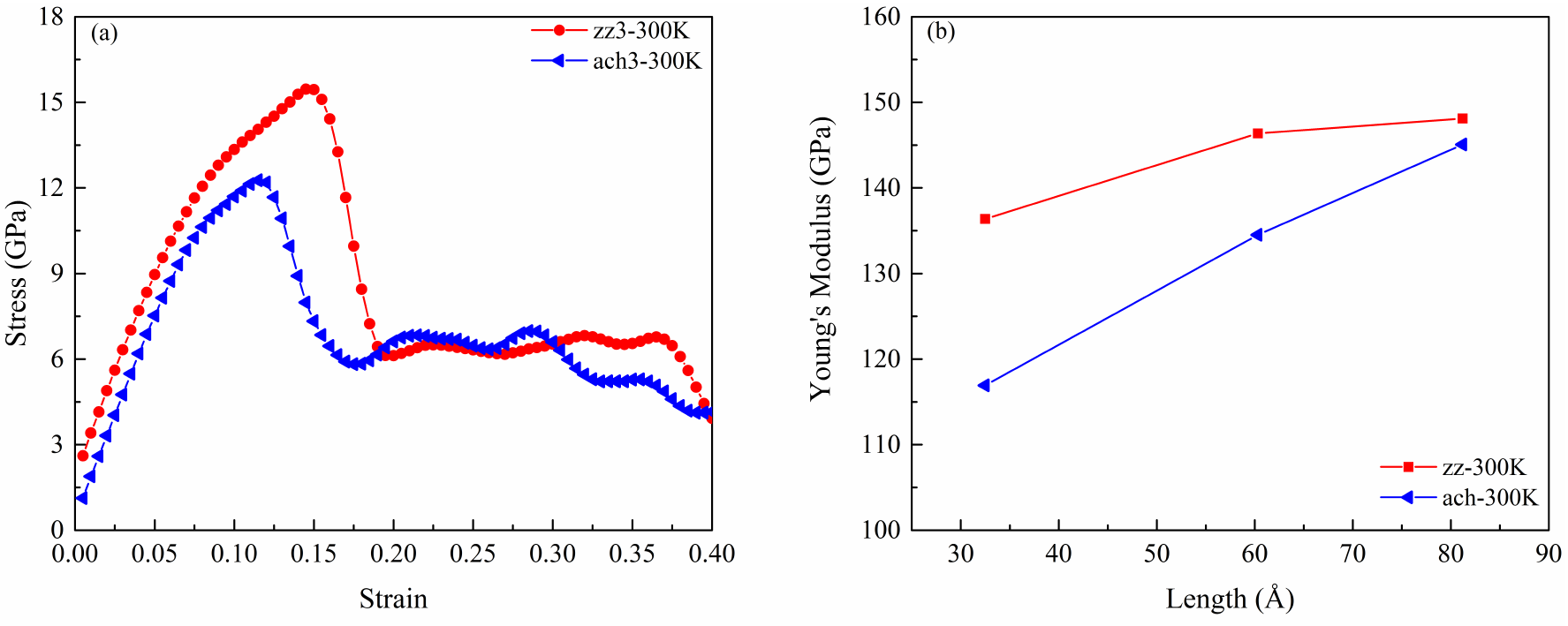}
  \caption{(a) Strain-stress relations of ach(zz)3 defect-free SNRs under tensile test. (b) Young's modulus for pristine SNRs as a function of length.}
  \label{fgr:YoungModulusSiliceneFig2}
\end{figure*}

Fig.~\ref{fgr:YoungModulusSiliceneFig2} (a) shows the strain-stress relations for the zz3 (red on line) and ach3 (blue on line) pristine SNRs under uniaxial tensile test. From the figure we can see the well defined elastic (linear) region at  small strain, followed by yielding until the Ultimate tensile strength is reached (maximum stress). Necking then continues until fracture (the latter one is not included here). Fig. \ref{SI-fgr:snapshots} in the SI shows the atomic configurations of a relaxed pristine silicene nanoribbon at different strains. YMs obtained for each defect-free ribbon are plotted in  Fig.~\ref{fgr:YoungModulusSiliceneFig2} (b). YM for both directions increases with ribbon length. This trend coincides with that reported in Ref. \cite{Jing2013}. Furthermore, zz-SNRs show larger values than those obtained for ach. Therefore, YM exhibits size  as well as chirality dependence. This behavior has also been found in graphene nanoribbons \cite{Zhao2009}.   The difference between the YM values with chirality occurs due to the orientation of the bonds along the tension direction. When a tensile load is applied along to the zz direction, {\it i. e.}, strain is applied perpendicularly to the ach direction, the stretched Si-Si bonds are inclined, (see Fig. \ref{fgr:YoungModulusSiliceneFig1}). Similar to graphene, bonds of this kind carry a part of the strain by changing the bond angles apart from increasing the bond length. Hence, deformations in the zz direction can be more supported. On the other hand, under ach deformation strain is applied in such a way that bonds parallel to the strain direction control deformation and their lengths will increase more easily than in the case of zz, leading to a smaller slope and, in consequence, a smaller YM  \cite{Dewapriya2013}. For the same reason, ach-SNRs show a stronger size effect. We can also observe that for both chiralities, YM approaches to the pristine sheet values around 100 \AA. 

\subsubsection{Defective nanoribbons.~~} 
\label{sec:defectivenrs}

\begin{figure}[h]
\centering
  \includegraphics[height=6cm]{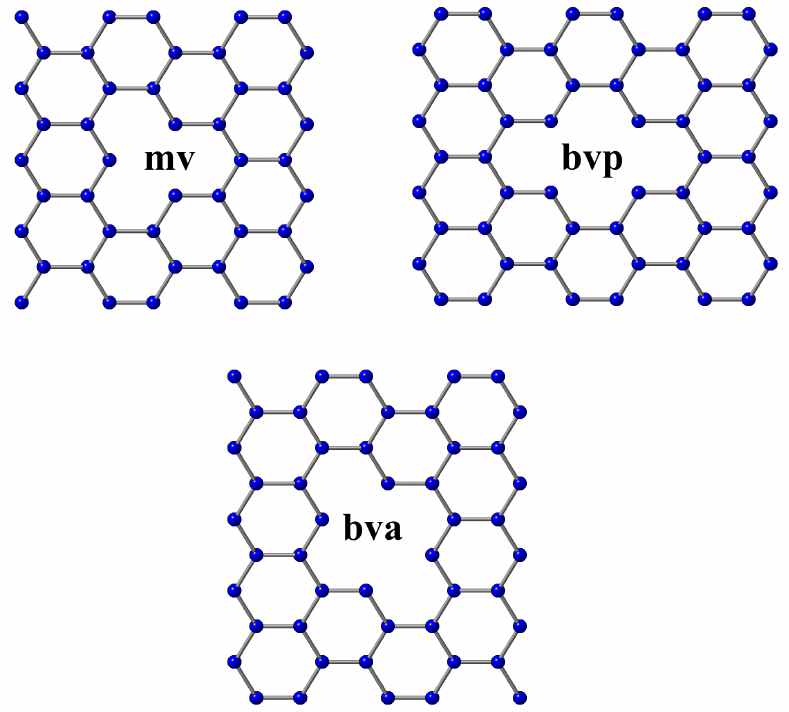}
  \caption{ Schematic representation of the vacancy-defects in SNRs. See text}
  \label{fgr:YoungModulusSiliceneFig3}
\end{figure}

Vacancy defects are created by removing atoms. We consider three kinds of them. Monovacancy (mv) in which one atom is removed,  and two types of bivacancies (bv) where  two adjacent atoms are eliminated. A bv is created by removing two adjacent atoms along the ach direction (bvp or parallel) or  the zz one (bva or angular). These defects are schematically illustrated in Fig.~\ref{fgr:YoungModulusSiliceneFig3}.  YM may be affected by the position of the lacking bonds since vacancies near to the edge are expected to make the ribbon more fragile. Three cases for the position are included. Vacancies  at the center of the ribbon  are labeled with "c" (mvc, bvac, and bvpc), while those close to the boundaries with "e" (mve, bvae, and bvpe). Next, combinations of one central and four externals are built. Notice that four externals are included (close to each corner of the NR) to have a symmetric ribbon, making the applied tensile stress along both chiral directions comparable. Since YM for pristine NRs depends on length, from now on results will be presented  in all cases normalized to the free-defective value and we will refer to them as normalized values or simply YM.

\paragraph{Mono- and bi- vacancy defects.~~} 

\begin{figure*}
 \centering
 \includegraphics[height=11cm]{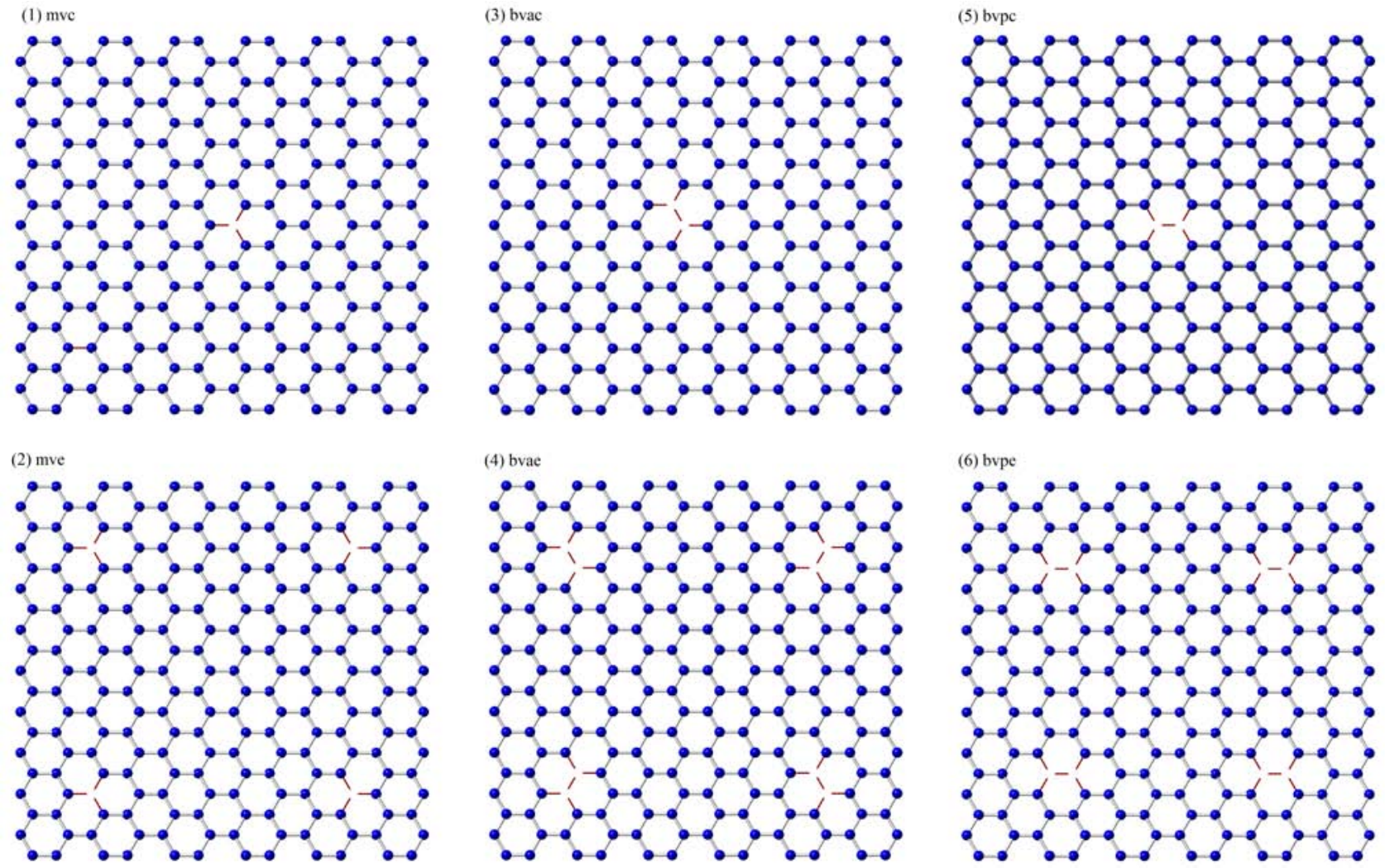}
 \caption{Schematic representation of the vacancy defects corresponding to the graphs shown in Fig.~\ref{fgr:YoungModulusSiliceneFig5}. Broken bonds are indicated with thin (red on line) lines, see text.}
 \label{fgr:YoungModulusSiliceneFig4}
\end{figure*}

Fig.~\ref{fgr:YoungModulusSiliceneFig4} shows the schematic representation of the central and external (close to the ribbon boundaries) mono- and bi- vacancy defects set into the studied ribbons. As explained before, external vacancies are situated symmetrically near the corners.

\begin{figure*}
 \centering
 \includegraphics[height=10cm]{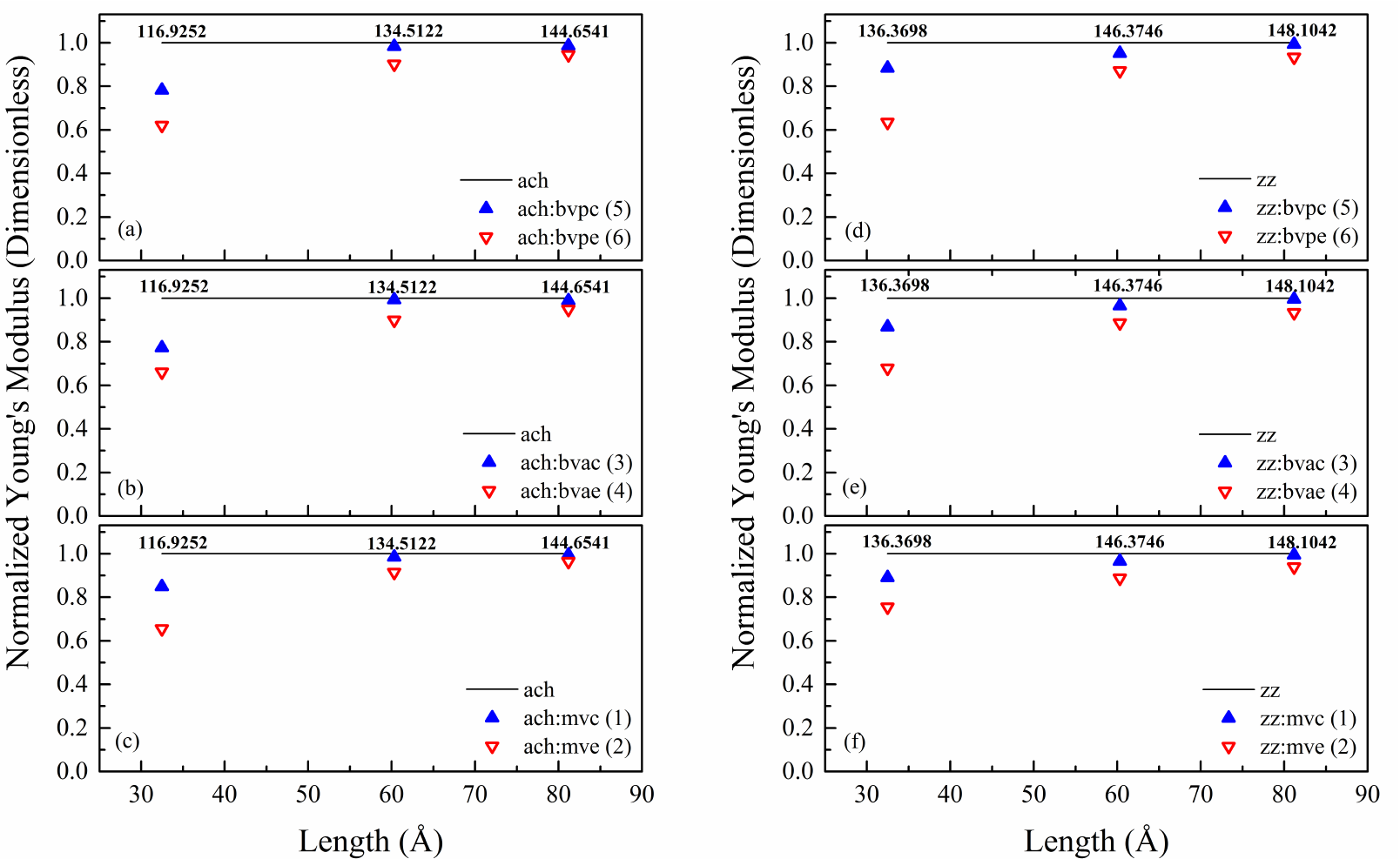}
 \caption{Young's modulus for defect-free and defective armchair and zig-zag SNRs {\it vs} length. Defects correspond to central (c) and externals (e) vacancies. On the top, numerical values correspond to the pristine SNR, in GPa. See text.}
 \label{fgr:YoungModulusSiliceneFig5}
\end{figure*}

In Fig.~\ref{fgr:YoungModulusSiliceneFig5} we  observe the effect on YM of type and position of the vacancy, as a function of length. On the top of each plot the value for the pristine nanoribbon is shown. As mentioned before, for a given length results are normalized to this value. Left (right) column corresponds to ach(zz) ribbons. Panels a, b and c correspond to bvp, bva and mv.  Numbers in the legend indicate the configuration of vacancies in the NRs shown in Fig.~\ref{fgr:YoungModulusSiliceneFig4}. It can be seen that  dependence on length is similar to the pristine case, {\it i.  e.}, the longer the ribbon the larger the Young's modulus.  We also observe that the change in YM due to vacancies will be negligible when the length approaches to 100 \AA. Le  {\it et al.} have reported a similar behavior of YM for a pristine sheet and that with one monovacancy in its center \cite{Le2015}.

We first analyze the ach case. Each panel shows effect of the vacancy position (see Fig.~\ref{fgr:YoungModulusSiliceneFig5}). Same trend with the length as in pristine NRs is observed for all vacancies.  YM for central vacancies (up triangles, blue on line) is larger than for externals due to the fact that the latter ones are four and then NRs have more missing bonds (5 for central {\it vs} 20 for externals bv,  while 3 {\it vs} 12 for mv, see Fig.~\ref{fgr:YoungModulusSiliceneFig4} for the structure). More missing bonds induces a larger deterioration of the structure, which causes bonds to carry less strain making the structure easier to deform. 
Regarding  the type, for bivacancies  (panels a and b in Fig.~\ref{fgr:YoungModulusSiliceneFig5}) YM does not show a significative difference since the number of missing bonds is the same for them (structures 3-6 in Fig.~\ref{fgr:YoungModulusSiliceneFig4}). The small difference could be explained in terms of the orientation of the missing bonds. While for the angular one (bvac) we have two missing bonds parallel to the deformation direction, for bvpc there is only one. Thus, bvpc has a larger relative value.

As for the zz direction (right column), the same behavior is observed in general. The main difference is  a larger separation between external (down triangle, red on line) and central (up triangle, blue on line)  bivacancies for the smallest ribbon (panels d and e in Fig.~\ref{fgr:YoungModulusSiliceneFig5}). Along this orientation,  the externals have a larger number of missing bonds (Fig.~\ref{fgr:YoungModulusSiliceneFig4}, structures 3-4 and 5-6). Therefore, YM are smaller than those for central vacancies.

\paragraph{Combinations of vacancy defects.~~}

\begin{figure*}
 \centering
 \includegraphics[height=17.5cm]{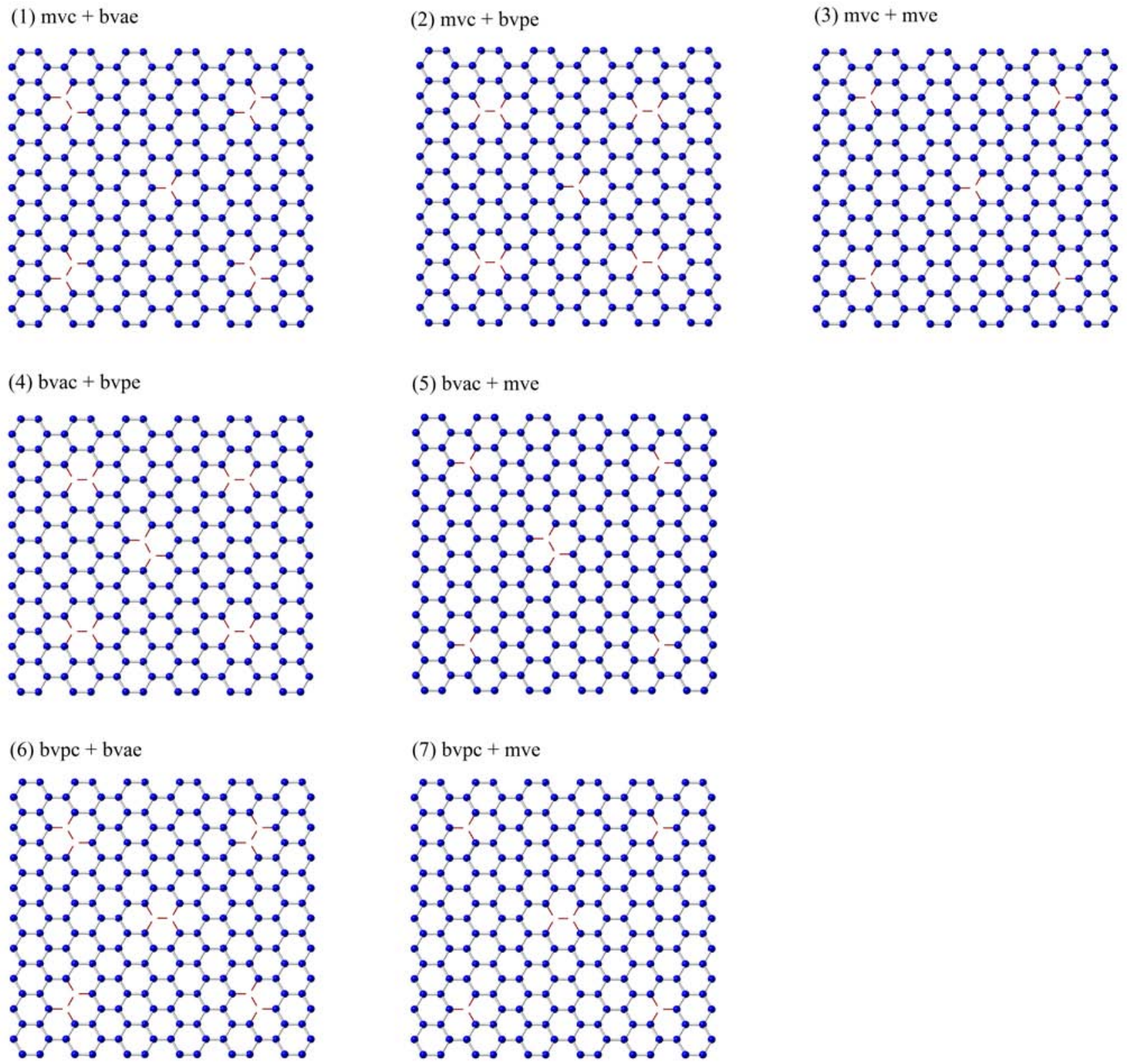}
 \caption{Schematical representation of the vacancy-defects corresponding to the graphs shown in Fig.~\ref{fgr:YoungModulusSiliceneFig7}.}
 \label{fgr:YoungModulusSiliceneFig6}
\end{figure*}

In Fig.~\ref{fgr:YoungModulusSiliceneFig6} we illustrate the different combinations of vacancy defects placed into the ribbons. 

\begin{figure*}
\centering
  \includegraphics[height=10.5cm]{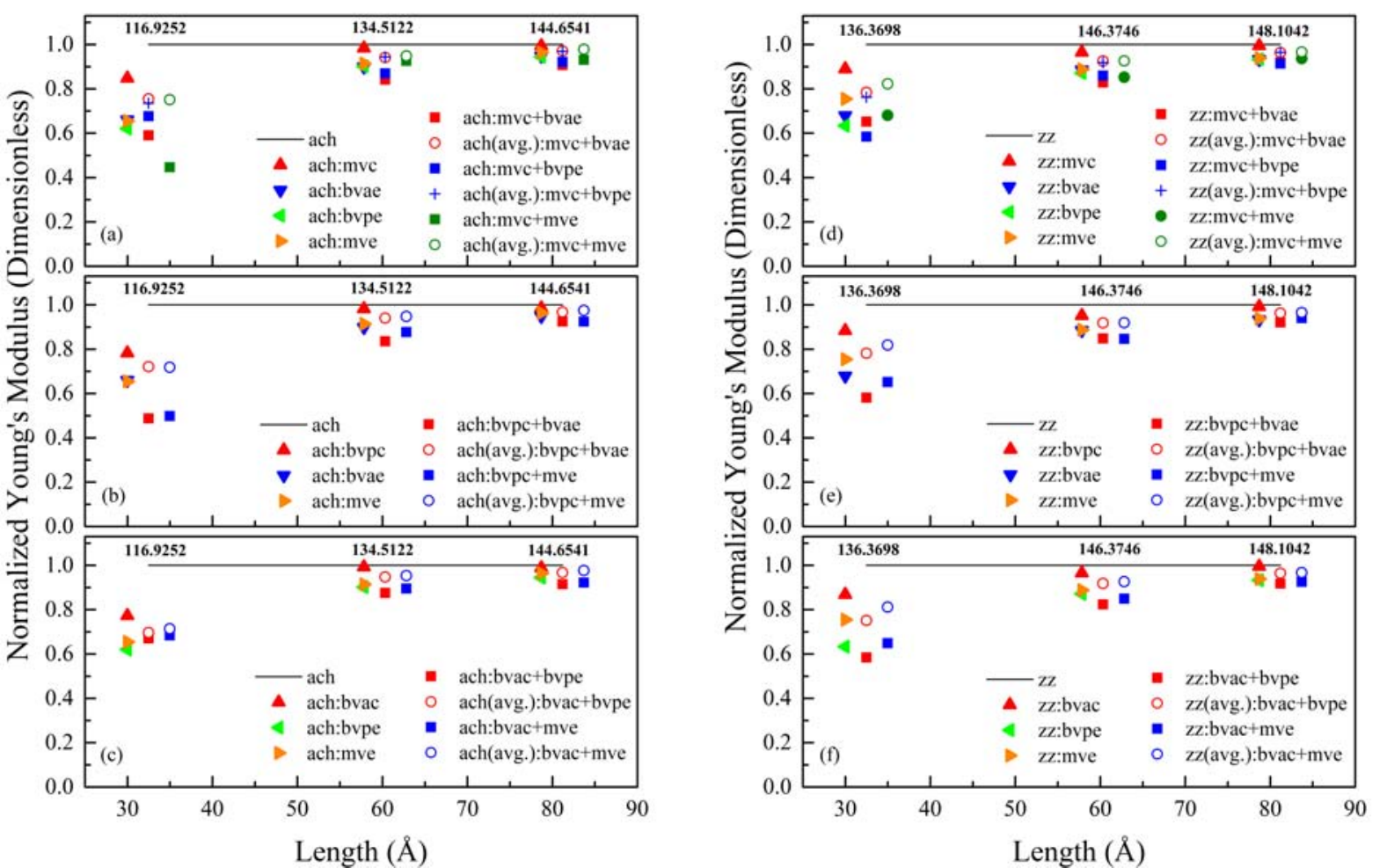}
  \caption{Young's modulus for defect-free and defective armchair and zig-zag SNRs {\it vs} length. Triangles correspond to non-combined vacancies, squared symbols represent  values obtained from the MD simulation while empty circles and crosses give the average ({\it avg}) of the corresponding non-combined NRs. On the top, numerical values correspond to the pristine SNR, in GPa. See text.}
  \label{fgr:YoungModulusSiliceneFig7}
\end{figure*}

In order to analyze the behavior of  normalized YM in the presence of combinations of vacancy-defects,  Fig.~\ref{fgr:YoungModulusSiliceneFig7} shows different kinds of YM values. As before, left (right) column corresponds to ach(zz) direction.
Each panel contains three columns of data for a given length. First column shows YM for NR with non-combined vacancies (shown in Fig. \ref{fgr:YoungModulusSiliceneFig5}). Second and third columns in turn give respectively YM as determined by MD (filled  squares) for the indicated combination and the average of the corresponding non-combined vacancies  (empty circles and crosses).  Panels a, b, and c  deal with  mv, bvp, bva located at the center of the NR. The overall behavior of bv follows the non-combined NRs, {\it i. e.}, the shorter the NR, the more sensitive is YM to length, see Sec. \ref{sec:defectivenrs}. In all cases, values given by the MD calculation are smaller than the average. The difference between them decreases as length increases, except  when the ach-NR has  an angular bivacancy at the center (bvac+bvpe and bvac+mve, panel c) since both combinations have the same number of missing bonds along the deformation direction. This is  overlooked for long NRs given their large number of bonds, thus both cases approach to the pristine case. The same number of missing bonds in the mentioned combination also causes that  the MD values are close to those of the non-combined NRs, opposite to the general trend of the rest for which the MD values are smaller. As for the zz direction, we observe a similar trend in the type of vacancy for corresponding panels (a and d, as well as b and e).  Thus, YM for combination of vacancies in general follows the same trend  with size as in the pristine case. However, YM is not a simple average of the values of the non-combined NRs , giving a complex relation between type and position of vacancies. Thus, tayloring of YM can be possible by means of a combination of vacancies of different types, for a given length and chirality.

\paragraph{Remarks.~~}

We give some general remarks. a) Relaxed structures with mve, show that vacancies do not coalesce into a multivacancy 
(see SI, Fig.~\ref{SI-fgr:snapshotsrelaxmv}). 
This behavior is preserved even in the presence of two nearby monovacancies, where vacancies do not move to form a bivacancy 
(see SI, 
FIg. ~\ref{SI-fgr:snapshotsrelaxmv}). These results are explained by the self-healing mechanism of vacancy-defects, which has already been studied in graphene and silicene.  In this process, the atoms surrounding the vacancy move towards the centre of the vacancy and the bonds are reoriented to close it. In the case of a monovacancy, the three dangling bonds from the atoms around the vacancy-defect form bonds that are stable and have the same length. For the bivacancy case, the atoms around the defect have four dangling bonds, which close the vacancy forming two Si-Si bonds \cite{Ozcccelik2013}. This reconstruction mechanism of the atoms avoids the coalescence of vacancies. b) Critical strains for SNRs under uniaxial tension are 19.5(17.5), 17.0(12.0), and 14.5(11.5) \%  in zz(ach) direction. As we can see, the critical strain decreases as length increases. These results show that the critical strain depends on the ribbons length. A similar behaviour has been observed in squared-shaped graphene-NRs (GNRs). On the one hand, zz-GNRs critical strain decreases with the ribbons width, and on the other critical strain for ach-GNRs slightly depends on the ribbons width  \cite{Chu2014} . Our results are similar to those reported for graphene as we are  also studying square-shaped structures, which by increasing its length also increase its width. c) We test the case where the vacancy concentration is the same in two NRs. In one, length is taken as $L$ (ach(zz)1) and the other has length $2L$. The first one has one mvc and the second one has four mv (see SI, Fig.~\ref{SI-fgr:snapshotsrelaxL}).  YM values obtained for the NR with length of $2L$ are 121.27 and 137.96 GPa  for {\it ach} and {\it zz} directions, respectively. While, YM for the NR with length of $L$ are 99.18 GPa for {\it ach}-direction and 121.40 GPa for {\it zz}-direction. These results show that not only the density of vacancy-defects are important as length increases, but also the number and position of vacancies strongly influence the value of YM. d) 
Fig.~\ref{SI-fgr:avegbuckling}a 
in SI shows the average buckling behaviour of a pristine SNR as function of the time step. It can be seen that the buckling amplitude varies with the time step, this dispersion is due to the displacement of the atoms in {\it x}-direction, which is compressed by exerting a tension in {\it y}-direction. Around 41 ps the structure has reached the necking region, and the buckling dispersion decreases as a result of the strain as expected, because with strain the structure tends to be planar. In Fig.~%\ref{SI-fgr:YoungModulusSiliceneESIBucklingFig2} 
\ref{SI-fgr:avegbuckling}b  of SI
the buckling for each atom confirms a qualitative change after necking.

 \subsection{Temperature effects}

\subsubsection{Mono- and bi- vacancy defects.~~} 

\begin{figure*}
\centering
  \includegraphics[height=10cm]{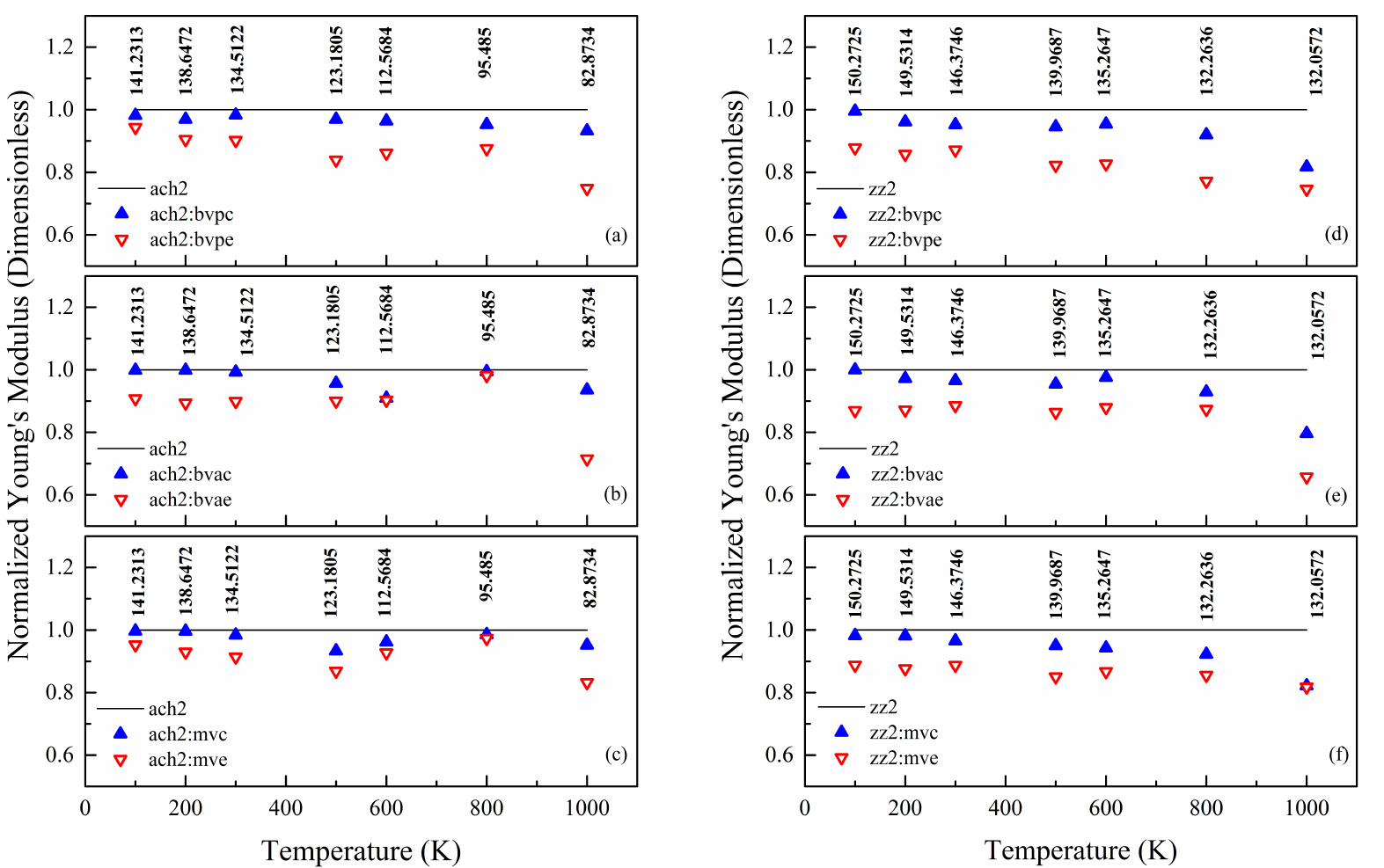}
  \caption{Young's modulus for defect-free and defective armchair and zig-zag SNRs {\it vs} Temperature. Defects correspond to central and close to the ribbon's boundaries vacancy defects (see Fig.~\ref{fgr:YoungModulusSiliceneFig4}). On the top, numerical values correspond to the pristine SNR, in GPa. See text.}
  \label{fgr:YoungModulusSiliceneFig8}
\end{figure*}

NRs of 60x60 \AA\ represent an intermediate size regime for our nanoribbons.  For this reason, we have chosen them to analyze the temperature effect.
Left (right) column of Fig.~\ref{fgr:YoungModulusSiliceneFig8} shows the normalized YM of defective ach(zz)2-SNRs at temperatures between 100 and 1000 K.  It is worth of mention that one atom misses during the calculation  at 1000 K. This might be interpreted as if one phase transition could take place. For this reason, this is the highest temperature we study. Panels a, b, and c correspond to bvp, bva and mv  types, respectively. On top of the panel,  the value of pristine NR is shown at the given temperature. As expected, the larger the temperature the smaller YM because at high temperatures the bond lengths are larger due to thermal motion of atoms, therefore the deformation resistance is lower. A similar behavior has been observed in pristine ach graphene, which even at 2400 K has 90\% its value at room temperature \cite{Zhao2010}. For silicene instead, the value has diminished to ~61 (90) \% of its value for ach(zz) at 1000 K. This is consistent with the known facts that graphene is stronger than silicene and pristine zz is in turn stronger than ach. However, while the pristine value follows a systematic  decrease, defective NRs show an overall complex behavior. YM decreases up to ~500 K as the pristine case, then it increases (depending on type of vacancy) and finally it drops. Central vacancies (up triangles, blue on line) have always larger relative values than the external (down triangle, red on line). As mentioned before, this is a consequence of the missing bonds. Some particular characteristics for vacancy types are described now. At some temperatures, YM does not depend on the type of vacancy as seen at 600 and 800 for the ach-bva whereas for ach-mv only at 800 K  (panels b and c, respectively). As for the zz case, this occurs  only at 1000 K for mv (panel f).  On the other hand, YM values for central and externals are similar around 800 K, but at higher temperatures they tend to separate.  However, an opposite behavior is observed for the zz-bv case (d). This shows that thermal motion causes atoms to arrange in a complex structure, giving rise to an alike temperature dependence but small differences determined by the vacancy type and chirality.

\subsubsection{Combinations of vacancy defects.~~} 

\begin{figure*}
\centering
  \includegraphics[height=10cm]{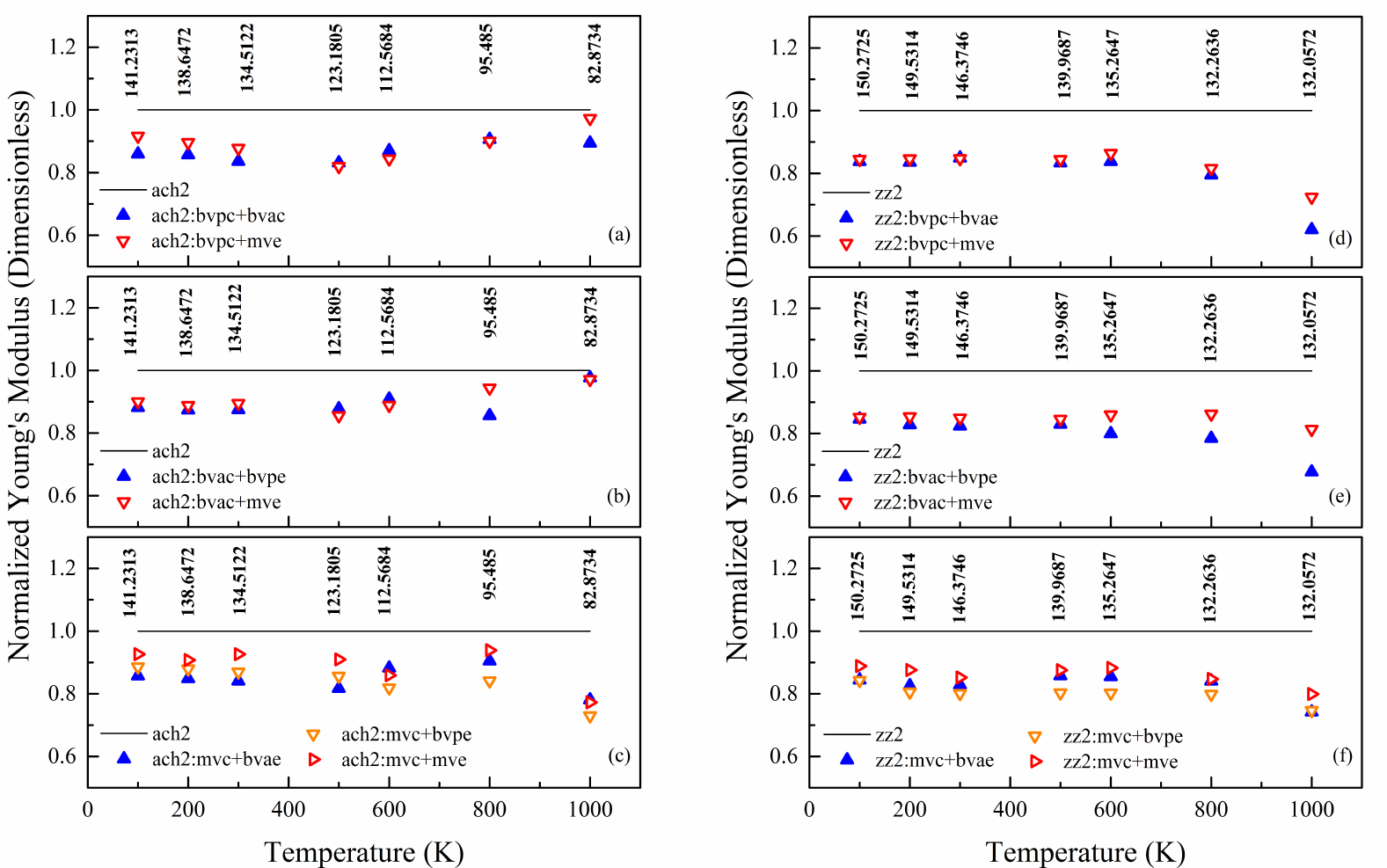}
  \caption{Young's modulus for defect-free and defective armchair and zig-zag SNRs {\it vs} length. Defects correspond to different combinations of vacancies (see Fig.~\ref{fgr:YoungModulusSiliceneFig6}). On the top, numerical values correspond to the pristine SNR, in GPa. See text.}
  \label{fgr:YoungModulusSiliceneFig9}
\end{figure*}

In Fig.~\ref{fgr:YoungModulusSiliceneFig9},  normalized YMs are plotted for different combinations of vacancy defects as a function of temperature.  As before, left (right) column corresponds to ach(zz)-NRs. The decrease of YM with increasing temperature remains, as in the previous case.  However, results do not show a strong dependence on the type of combined vacancies up to 600 K. At higher temperatures, differences respect to the pristine case are significant only for combinations with mvc (panel c). In contrast, all zz-NRs (right panels) show a decrease and a clear difference for the combinations is observed at high temperatures. This is due to the stronger bonds along the zz direction such that atoms rearrange less easily compared to the ach. Therefore, chirality plays an important role in this case. A more detailed study of the structural behavior is in progress.  

\section{Conclusions}
\label{sec:Concl}
We have performed molecular dynamics simulations to study the Young's modulus of silicene nanoribbons, with and without vacancy defects, as a function of length and temperature. At room temperature, the YM obtained for pristine nanoribbons nonlinearly increases with ribbon length. YM values for zz-SNRs are larger than those obtained for ach-SNRs. The difference arises due to the different type of bonds along the tension direction, being the zz one the stronger as reported before. Three cases of vacancies (mono-, parallel and angular vacancies) were analyzed and, in all the studied cases, YM increases with length for the given number of vacancies. Thus, mechanical properties of defective SNRs exhibit a size dependence. As length increases,  YM tends to the corresponding pristine value, which means that the vacancy defects in the SNRs considered in this work do not affect the YM as the size approaches to 100 \AA.\ Combinations of vacancies give a rich and complex behavior. Values obtained by numerical calculations do not coincide with the average of the YMs of non-combined vacancies.  As for the temperature, calculations were performed for an intermediate length, up to 1000 K.  YM for the pristine NRs nonlinearly decrease with increasing temperature. When vacancies are included, it shows a similar nonlinear-dependence regardless the type of vacancy, but  different to the pristine case. However, some combinations show particular features which depend significantly on the temperature.
Thus, we have shown that tailoring of Young's modulus can be done by means of a combination of vacancies of different types, for a given length and chirality. Our results can be helpful for the design and performance of devices based on these nano materials.

A main subject of further investigations, it is to find the Young Modulus of silicene on a substrate. As we mentioned before, silicene could form different structures on metal surfaces \cite{Kaltsas2014, Gao2014,Quhe2014}. Since Ag and Si atoms do not easily form an alloy, Silver has been considered the appropriate substrate for the deposition of silicene \cite{Enriquez2012, Feng2012}. However, to study the YM of silicene on a Ag-substrate we have to consider not only the strains that the substrate induce to the material due to the lattice mismatch,  but also the interactions of silicon atoms with the substrate, the effect of the substrate temperature and the hybridization between silicon and silver atoms \cite{Jamgotchian2012, Yuan2014}. All these factors can strongly influence the value of the YM, because they should modify the strain tensor components giving rise to a different value. Thus, YM values obtained here are to be considered lower bounds.

\section*{Acknowledgement}
Partially supported by CONACYT and VIEP-BUAP, M\'exico.

%%%END OF MAIN TEXT%%%

%The \balance command can be used to balance the columns on the final page if desired. It should be placed anywhere within the first column of the last page.

%\balance

%If notes are included in your references you can change the title from 'References' to 'Notes and references' using the following command:
%\renewcommand\refname{Notes and references}

%%%REFERENCES%%%
\bibliography{bibfile} %You need to replace "rsc" on this line with the name of your .bib file
\bibliographystyle{rsc} %the RSC's .bst file

\appendix
\section{Supplementary Information (SI)}

\begin{figure}[h]
\centering
%\subfigure[Front view \label{fgr:YoungModulusSiliceneESIRelaxaFig1}]
(a) \\
{\includegraphics[height=6cm]{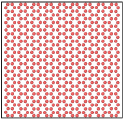}}\hspace{10mm}
%\subfigure[Lateral view \label{fgr:YoungModulusSiliceneESIRelaxbFig2}]
\\ (b) \\
{\includegraphics[height=6cm]{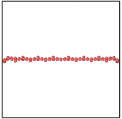}}\vspace{10mm}
\caption{ Snapshots of atomic configurations of relaxed pristine silicene nanoribbon of 60 \AA\  side at 300K.} 
%\label{fgr:snapshotsrelax}
\label{SI-fgr:snapshotsrelax}
\end{figure}

\noindent
Initial conditions for atoms are the positions of ideal low-buckled silicene NR. After relaxation, although atoms seen to have ideal positions on the plane (Fig.~\ref{SI-fgr:snapshotsrelax}a), buckling is not uniform through the structure as 
Fig.~\ref{SI-fgr:snapshotsrelax}b.

\begin{figure}[h]
\centering
%\subfigure[0\% strain \label{fgr:YoungModulusSiliceneESIFig1}]
(a) \\
{\includegraphics[height=4cm]{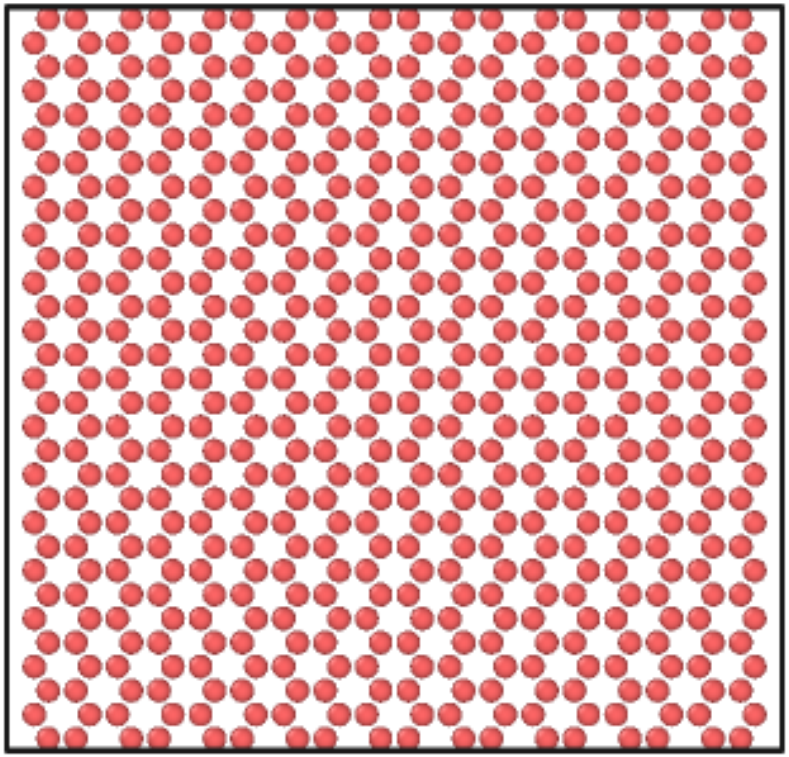}}%\hspace{10mm}
%\subfigure[16\% strain \label{fgr:YoungModulusSiliceneESIFig2}]
\\ (b) \\
{\includegraphics[height=4.55cm]{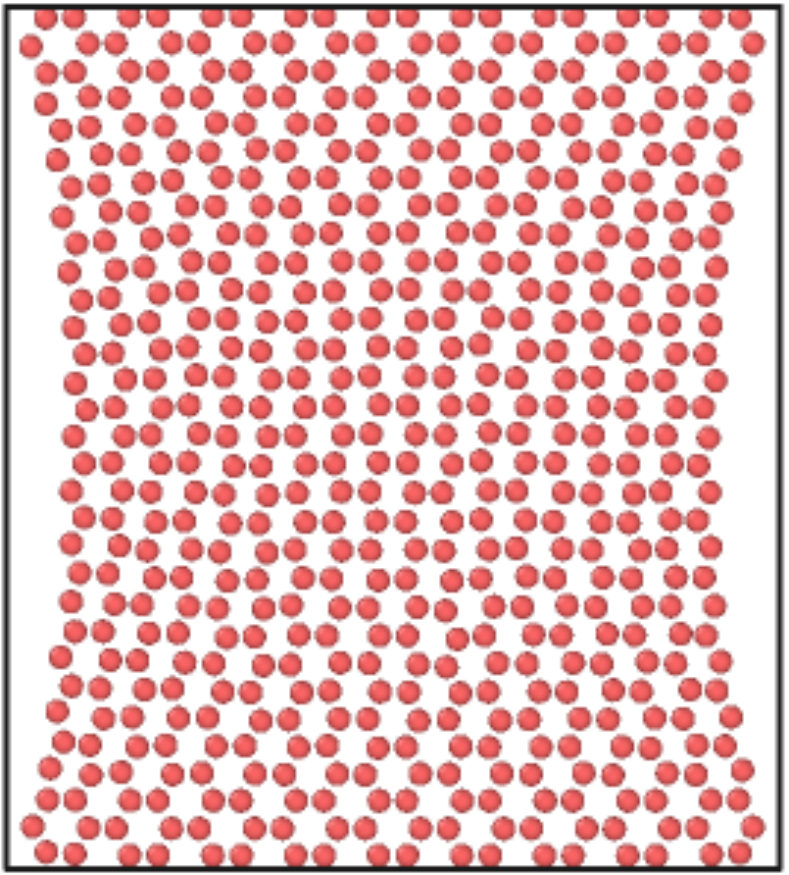}}%\vspace{10mm}
%\subfigure[30\% strain \label{fgr:YoungModulusSiliceneESIFig3}]
\\ (c) \\
{ \includegraphics[height=5cm]{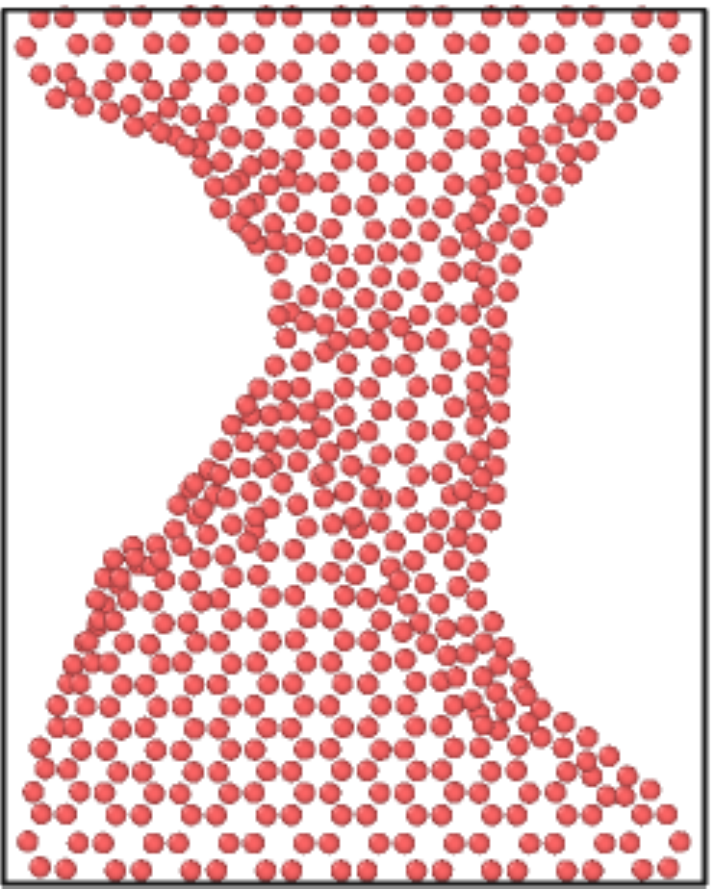}}%\hspace{17mm}
%\subfigure[45\% strain \label{fgr:YoungModulusSiliceneESIFig4}]
\\ (d) \\
{\includegraphics[height=5.5cm]{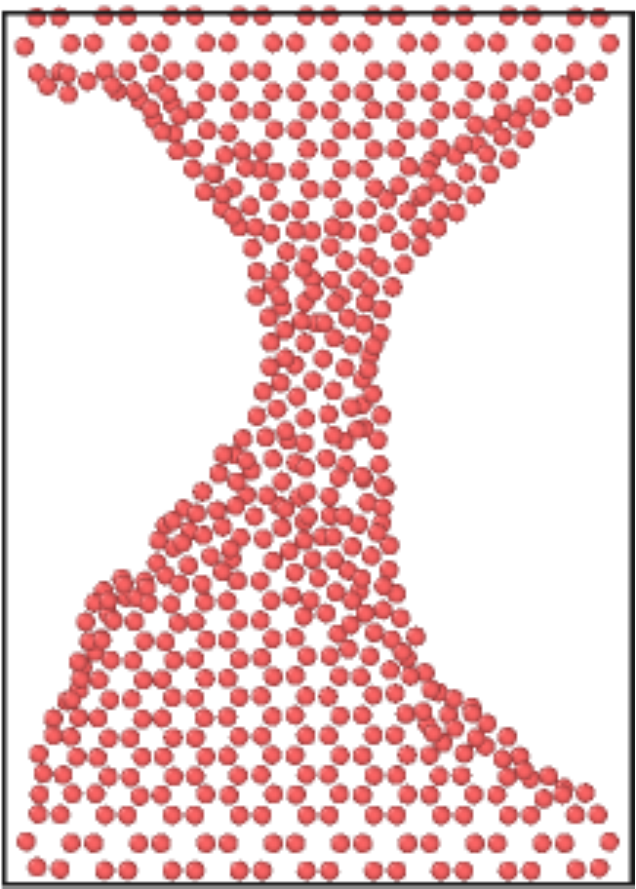}}%\vspace{10mm}
\caption{ Snapshots of atomic configurations of pristine silicene nanoribbon of 60 \AA\  side, at different strains along zig-zag direction.} \label{SI-fgr:snapshots}
\end{figure}

In FIg. \ref{SI-fgr:snapshots} 
we show snapshots that represent different points on the stress-strain curve. 
Fig.~\ref{SI-fgr:snapshots}a 
shows the relaxed silicene structure, corresponding to zero deformation. 
Fig.~\ref{SI-fgr:snapshots}b
corresponds to the ultimate tensile strength, just before necking occurs. SNRs considered in this study show ductile behavior due to the formation of the necking before reaching its breaking point . 
Fig.~\ref{SI-fgr:snapshots}c 
and Fig.~\ref{SI-fgr:snapshots}d 
present the necking process of the silicene structure. 
The critical strains of SNRs are at 14.5\% and 11.5\% for zz and ach, respectively. As we can see,  SNR's critical strains are smaller than those obtained for silicene sheets (19.5\% for zz and 15.5\% for ach) \cite{Jing2013}. 

\clearpage

\section{Vacancy self-healing}

\begin{figure}[htb!]
\centering
%\subfigure[Monovacancies close to the ribbon's boundaries (mve)  \label{fgr:YoungModulusSiliceneESIRelaxmveFig1}]
(a) \\
{\includegraphics[height=5.9cm]{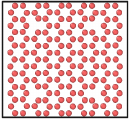}}%\hspace{10mm}
%\subfigure[Monovacancies close to each other  \label{fgr:YoungModulusSiliceneESIRelax2mvcFig2}]
\\ (b) \\
{\includegraphics[height=6cm]{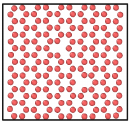}}%\vspace{10mm}
\caption{ Snapshots of atomic configurations of relaxed defective SNRs of 32 \AA\  side.} \label{SI-fgr:snapshotsrelaxmv}
\end{figure}

\noindent
Moonovacancy-defects do not coalesce into a multivacancy in the relaxed structures.

\section{Density dependence}

\begin{figure}[htb!]
\centering
%\subfigure[Defective NR of length L \label{fgr:YoungModulusSiliceneESIRelaxLmvFig1}]
\vspace{10mm}
(a) \\
{\includegraphics[height=6cm]{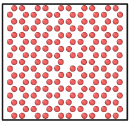}}%\hspace{10mm}
%\subfigure[Defective NR of length 2L \label{fgr:YoungModulusSiliceneESIRelax2LmvFig2}]
\\ (b) \\
{\includegraphics[height=6.5cm]{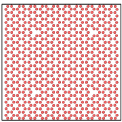}}%\vspace{10mm}
\caption{ Snapshots of atomic configurations of relaxed defective SNRs, whose length varies from L (32 \AA) to 2L .} \label{SI-fgr:snapshotsrelaxL}
\end{figure}

\noindent
Both SNRs show have the same density of defects, but different YM values.

\section{Buckling}

\begin{figure}[htb!]
\centering
%\subfigure[Average buckling \label{fgr:YoungModulusSiliceneESIABucklingFig1}]
(a) \\
{\includegraphics[height=6cm]{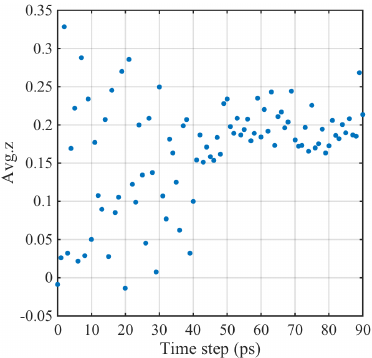}}\hspace{10mm}
%\subfigure[Buckling position \label{fgr:YoungModulusSiliceneESIBucklingFig2}]
\\ (b) \\
{\includegraphics[height=6cm]{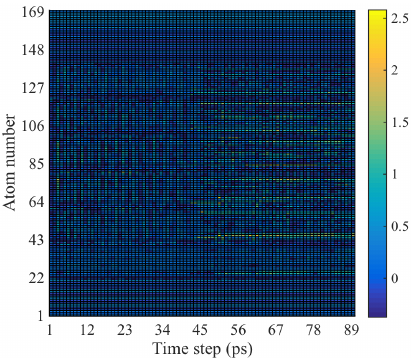}}\vspace{10mm}
\caption{Buckling behaviour of the zz-SNRs under uniaxial tension.(a) shows the average value of $|z(t)|-z(t=0)$, over the total number of atoms. (b) Buckling position for each atom defined as $|z(t)|-z(t=0)$.}\label{SI-fgr:avegbuckling}
\end{figure}

\noindent
Notice that the average buckling amplitude changes importantly during the lineal regime and the plastic region. Around 40 ps the  ultimate tensile strength (UTS) is reached. After that point the structure is in the necking region and the buckling dispersion decreases as a result of the applied tension, indicating that the structure tends to be planar. Atomic buckling, 
Fig. \ref{SI-fgr:avegbuckling}, also shows a change of regime at 40 ps.

\end{document}